# Eco-engineering controls vegetation trends in southwest China karst

Xuemei Zhang [a, b, c], Yuemin Yue [a, b], Xiaowei Tong [d], Kelin Wang [a, b, *], Xiangkun Qi [a, b], Chuxiong Deng [e], Martin Brandt [d]

[a] Key Laboratory of Agro-ecological Processes in Subtropical Region, Institute of Subtropical Agriculture, Chinese Academy of Sciences, Changsha 410125, China

[b] Huanjiang Observation and Research Station for Karst Eco-systems, Huanjiang 547100, China

[c] University of Chinese Academy of Science, Beijing 100049, China

[d] Department of Geosciences and Natural Resource Management, University of Copenhagen, Copenhagen 1350, Denmark

[e] College of Resources and Environmental Sciences, Hunan Normal University, Changsha 410125, China

*Corresponding authors:

Kelin Wang (Email: kelin@isa.ac.cn; Tel.: 86-0731-84615201; Fax: 86-731-84612685)

Key Laboratory of Agro-ecological Processes in Subtropical Region, Institute of Subtropical Agriculture, Chinese Academy of Sciences, Changsha 410125, China

## Abstract

The karst area in Yunnan-Guangxi-Guizhou region in southwest China is known for widespread rocky desertification but several studies report a greening trend since the year 2000. While the start of the greening trend seems to match with the implementation of ecological conservation projects, no statistical evidence on a relationship between vegetation greening and eco-engineering exists. Moreover, dominant factors influencing the spatial patterns of vegetation trends have rarely been investigated. Here we use six comprehensive factors representing the natural conditions and human activities of the study area, and several statistical models consistently show that eco-engineering explains large parts of the positive vegetation trends in the karst areas, while negative vegetation trends in non-karst areas of Yunnan were related with a decrease in rainfall. We further show that the interaction of eco-engineering with other factors leads to a heterogeneous pattern of different vegetation trends. Knowing and understanding these patterns is crucial when planning ecological restoration, especially in diverse landscapes like China karst and the methods can be reused in other restoration areas.

## 1. Introduction

Vegetation is an important part of terrestrial ecosystems and plays a critical role in regulating the carbon balance (Piao et al. 2009, Li et al. 2012, Hu et al. 2018), but is highly sensitive to climate change and human activities (Zhao et al. 2019a). Remote sensing can provide repeated observations to gain insights into the dynamics of vegetation at large scales (Guo et al. 2017). Numerous studies have reported greening trends, interpreted as increases in vegetation cover (Piao et al. 2015, Brandt et al. 2018, Song et al. 2018, Zhao et al. 2020a, Hong et al. 2020). Especially in southwest Chinese karst area, characterized by a highly heterogeneous landscape (Guo et al. 2020), a general greening trend was reported based on time series of satellite observations and model predictions (Hou et al. 2015, Tong et al. 2016, Tong et al. 2017, Brandt et al. 2018, Tong et al. 2018, Xu et al. 2019, Liu et al. 2020). The reliable detection and attribution of factors impacting on vegetation trends are a prerequisite for the development of strategies for the sustainable management of ecosystems (Piao et al. 2015), and therefore the main drivers of large scale increases in vegetation cover have been in the scientific spotlight over the past decades (Peng et al. 2011, Tian et al. 2013, Piao et al. 2015, Sun et al. 2015, Tong et al. 2018, Piao et al. 2020).

Several studies have revealed that global warming stimulates vegetation growth by extending the growing season and promoting summer photosynthesis, particularly in regions where water is not a limiting factor (Piao et al. 2007, Piao et al. 2008), but also precipitation can play a crucial role on vegetation change (Zhao et al., 2020a). Therefore, it is critical to quantify the impact of different climate factors on vegetation trends in different regions, especially in karst and non-karst areas. The impact of soil, topography and geological factors on vegetation changes in karst areas has been studied (Liu et al. 2020, Qiao et al. 2020), but the factors are usually treated as single factors, and their combined effects are rarely considered. Furthermore, most studies focus on small sample size (Hu et al. 2018, Wang et al. 2018, Zhang et al. 2019) without large-scale application.

The rapid expansion of human population and increasing demand for natural resources, such as agricultural land and forest products (Brandt et al. 2017, Qiu et al. 2020), threatens the eco-environment (Li et al. 2017). Many studies have discussed that human disturbances are threatening biodiversity and lead to land degradation (Martínez-Ramos et al. 2016, Nguyen and Liou 2019). However, several studies have also emphasized the potential role of increased carbon sequestration from adequate forest management (Tong et al. 2020) and vegetation recovery in rural areas arising from agricultural abandonment and the movement of the rural population to cities (Piao et al. 2015, Hu et al. 2018).

Recently, large parts of the greening trend are attributed to ecological conservation projects like the Grain for Green program (Brandt et al. 2018, Tong et al. 2018, Zhao et al. 2019b) aiming at recovering ecosystems, alleviate poverty (Zhang et al. 2017, Wang et al. 2019b, Zhou et al., 2020b) and improving ecosystem services (Qiu et al. 2020, Zhou et al. 2020a, Liu et al. 2008, Wang et al., 2020). Methods looking into land use change (Liao et al. 2018, Liu et al. 2020), rocky desertification control (Jiang et al. 2014) and vegetation improvement (Qi et al. 2013) were used to indicate the effectiveness of ecological engineering. However, although the impact of eco-engineering on vegetation trends seems to be visually clear (Tong et al. 2017, Brandt et al. 2018), the individual and interacting contributions of climate conditions and human activities to vegetation trends need to be quantified in a statistical way.

Complex environmental processes interact with each other and factors are typically not independent (Zhao et al. 2017, Wang et al. 2019). Previous studies mainly focused on a single factor, but it is important to study a variety of factors and the interaction between them. The combined impact of poverty transition and other factors on vegetation change in South China karst studied by Zhao et al. (2020b) has indicated that a synergetic impact on vegetation recovery between ecological engineering and poverty alleviation exists. However, the combined impact of ecological engineering and soil, socio-economical factors and interference of human activities on vegetation change in different project regions has not yet been studied, which is crucial when planning ecological restoration.

In this study, the impact of natural (climatic conditions, soil properties, geological factors and river density) and human factors (road density, GDP, land use types, population density, capital investment and afforestation area reflecting eco-engineering) on vegetation trends in China karst was study. Geo-detector model (GDM) (Xu et al., 2014, Wang and Xu, 2018) was used to detect the contributions of the factors on vegetation trends, and to quantify the interactions between ecological engineering and the other factors. Moreover, Geographical Weighted Regression model (GWR) was used to investigate the spatial relationships between the factors and vegetation trends (Hou et al., 2020).

We studied vegetation trends in different project regions, reflecting different geological settings. The objectives of this study are to identify the dominant factors impacting vegetation trends in different project regions of southwest China and to study which factors can enhance the impact of eco-engineering on vegetation trends. This work is important for decision-makers and stakeholders for planning, adjusting and assessing ecological engineering projects.

## 2. Materials and Methods

### 2.1 Study area

The study area covers the Yunnan, Guangxi and Guizhou provinces of which 35% are karst, dominated by pure carbonate bedrock (Tong et al. 2017). The area is one of the largest contiguous karst areas in the world, with more than 30 million people largely living under poverty (Zhang et al. 2017a). A mild and humid subtropical monsoon climate dominates the region while the mean annual rainfall is 1021 mm. Subtropical evergreen and deciduous broad-leaved forests are the major vegetation types, while farmlands cover only 10% (Wang et al. 2007, Tong et al. 2017). The terrain is high on the Yunnan Plateau and low in the southeast coastal area of Guangxi. Topography, lithology and geological conditions can be used to group the study area into eight regions, which are important to consider when planning eco-engineering projects (Yuan 2014) (Fig. 1).

The study area includes 294 counties with an average size of 2,709 km$^2$. The population is relatively poor, and the density of 126 persons per km$^2$ is above the theoretical carrying capacity (100 persons per km$^2$) in this karst area (SFAB 2018). Due to prolonged human activities, the karst area has experienced severe rocky desertification (Tong et al. 2016). Since the late 1990s, ecological engineering projects like the Grain for Green project required most sloping farmlands to be converted to forests or grasslands (Liu et al. 2008, Tong et al. 2017, SFAB 2018, Hu et al. 2018a, Hu et al. 2018b, Yue et al. 2020). Furthermore, the Karst Rocky Desertification Comprehensive Control Project running from 2005-2016, has decreased rocky desertification area on 289.2 km² (SFAB 2012, 2018).

*2.2 Datasets*

We use the GIMMS-3g Normalized Difference Vegetation Index (NDVI) time series derived from Advanced Very High Resolution Radiometer (AVHRR) covering the period from 1982 to 2016, available at 8 km spatial resolution (Pinzon and Tucker 2014). The dataset provides two images per month, and we selected the pixels with highest value to form a monthly time series. We then aggregated values from April to November to estimate the vegetation production over the growing season, also termed growing season NDVI (GSN) for each year (Tong et al. 2016). Eco-engineering projects started in 2000, and we use this year to split the time series into two periods (1982-2000 and 2001-2016). In spite of low spatial resolution, the long time series of the dataset is crucial to compare spatial and temporal vegetation changes before and after the implementation of ecological engineering (Hou et al., 2015, Tong et al., 2017).

The factors that impact vegetation growth during the conservation period (2001-2016) include natural and human factors, all factors are related to water, soil, heat, geological background and human activities. Natural factors include annually accumulated temperature >10°C, mean annual temperature (2001-2015), annually accumulated temperature >0°C, mean annual precipitation (2001-2015), aridity, sand content (%), silt content (%) and clay content (%), soil nutrients, geological settings (DEM) (Fig. 1) and hydrologic condition. Human factors include land use types, population density, GDP, road density, and ecological engineering (capita investment and afforestation area) (Table 1). All raster datasets were downloaded from the Research Center for Eco-Environmental Sciences, Chinese Academy of Sciences (http://www.resdc.cn/). Statistical data on the Grain to Green Program, which was used to represent the ecological engineering, were available at county scale and include afforestation areas (e.g. areas for mountain closure, afforestation, and cropland conversion) and funding (e.g. money allocated for grain and seeding) from 2001 to 2015, and were provided by the Forestry Bureau of the Yunnan, Guizhou, and Guangxi Provinces (Tong et al. 2017). Land use types in 2005, 2010 and 2015 were obtained from the Institute of Remote Sensing and Digital Earth, Chinese Academic of Science (http://www.ceode.cas.cn/sjyhfw/). Land use types were quantified as categorical variables using the Land use degree comprehensive index (LDCI) (Guo et al. 2018), the equation is Eq.1:

$$\text{LDCI}_a = 100 \times \sum_{i=1}^{n} A_i \times C_i$$

(1)

Where $\text{LDCI}_a$ is the land use degree index; $A_i$ is the land use classification index and the quantitative values are 4, 3, 2, 2, 2, 1 for artificial area, farmland, woodland, wetland, grassland and unused land; $C_i$ is the area percentage of different land use types in one unit. We then used the average LDCI of the three year 2005, 2010 and 2015. All data were averaged per county and then normalized from 0 to 1, to ensure the comparability of different data sources at different spatial scales and with different units (Qiu et al. 2020). All used datasets and their sources are listed in Table 1.

*2.3 Research methods*

The study uses long time series remote sensing data to infer vegetation trends (GSN slope) using linear trend analysis. We compared spatial patterns in GSN slope based on spatial autocorrelation analysis before and after the implementation of eco-engineering. Geographical Detector Model (GDM) (Wang and Xu 2017) was used to quantify the relative importance of drivers and their interactions with GSN slope. In addition, our study uses the Geographical Weight Regression model (GWR) (Brunsdon et al. 1996) to explore spatial non-stationarity correlations between drivers and GSN slope, as well as the sensitivity of GSN slopes to different drivers (expressed by the regression slope). The research methods of this study are summarized in Fig. 2 and consist of three main steps: (1) the calculation of linear trends in vegetation cover (GSN slope) for pre- and post- project periods; (2) the grouping of factors to comprehensive factors using principal component analysis (PCA); (3) the identification of drivers of GSN trends using GDM and GWR.

*2.3.1 Spatial pattern of vegetation trends*

The slope derived from linear trend analysis shows the annual change rate of vegetation greenness in GSN units. We used long time series remote sensing data to infer vegetation trends (GSN slope) from the linear trend analysis. Eco-engineering projects started in 2000, and we use this year to split the study period (1982-2016) into a reference period (1982-2000) and conservation period (2001-2016) following Tong et al. (2017). The mean GSN trends were averaged per county to be comparable with the other datasets. The equation to calculate the GSN slope is in Eq. S1. We further used the global spatial autocorrelation Moran's I ($I_g$) (Anselin et al., 2006) and local spatial autocorrelation Moran's I ($I_l$) (Anselin 1995) to find spatial patterns and spatial clusters of GSN increase/decrease (Supplementary materials).

*2.3.2 Extraction of comprehensive factors using principal component analysis*

In order to eliminate multicollinearity between the 20 factors, the Principal

Component Analysis (PCA) was used to extract principal components (PCs), which we refer to as comprehensive factors (Gao et al. 2006, Zhang and Dong 2012). The cumulative contribution rate of the first 6 PCs (PC1-PC6) (Huang et al. 2014, Xie et al. 2016, Zhang et al. 2017b) is 80.01% (Table S3). The PCA results are shown in Table S3, according to which the PCs were classified into 6 comprehensive factors and their detailed description is as follows: The contribution of the first PC is 24.93% which is mainly from climatic variables; therefore, PC1 is referred to as climate conditions. The second PC contributes 13.87% mainly from population density and land use degree comprehensive index, and is therefore referred to as human activities. The third PC contributes 12.15% mainly from sand and clay content, and is therefore referred to as soil texture. The fourth PC contributes 11.51% mainly from afforestation area and investment in ecological engineering funds; therefore, it is referred to as ecological engineering. The fifth PC is referred to as soil nutrients, having most of its loading from soil organic carbon content under different soil conditions and organic nitrogen storage, and it contributes 9.55%. The contribution of the sixth PC is 8.00%, and it has most of its loading from traffic density and GDP, and is therefore referred to as socio-economic conditions.

*2.3.3 Geographical Detector Model (GDM)*

GDM is a statistic method to detect the relative importance of individual factors (and their interactions) to response variables by testing for spatial correspondence between the variables (Wang and Xu 2017). We apply GDM to quantify the spatial agreement between environmental factors and GSN slopes.

The degree of spatial correspondence between GSN slope and an environmental factor is measured by the power of determinant (PD), which is calculated as follows:

$$P_{D,H} = 1 - \frac{\sum_{i=1}^{L} n_i Var_i}{nVar}$$

(2)

where D represents the environment factors; $i = 1, 2, \cdots, L$ denotes that the study area is stratified into $L$ strata (Wang and Hu 2005) according to the spatial patterns of an environmental determinant; denotes the number of counties in the study areas; H refers to the GSN slope; $P_{D,H}$ is the power of determinant of the environmental factor D on H; $Var_i$ and $Var$ denote the variance of GSN slopes in each stratum and over the entire study area, respectively. PD ranges from 0 to 1, with higher PD values reflecting stronger explanatory power of the variables. GDM can also test if the interaction between two factors impacts on the explanatory power of GSN slope (in this study interaction is symbolized by ∩).

The study uses the GSN slope during the conservation period (2001-2016) as the dependent variable and the six PCs (comprehensive factors) are the explaining variables. All six comprehensive factors are continuous variables, and thus had to be stratified though discretization before they could be used in the GDM (Wang and Xu 2017, Wang and Xu 2017, Wang et al. 2010, Ge et al. 2017, Wang et al. 2017a). In order to avoid subjectivity and randomness in the process of discretization, this study used the natural break classification method, which performs an automatic unsupervised segmentation of the comprehensive factors to receive the breakpoints between the segments (Xu and Zhang 2014). Finally, the optimal number of classes for each comprehensive factor was determined by the PD value, which expresses the degree of spatial correspondence between the GSN slope and the comprehensive factors (Cao et al. 2013, Ju et al. 2016). Fig. S1 shows which number of classes corresponds with the highest PD value of each comprehensive factors, and the chosen number of classes for each comprehensive factor (from PC1 to PC6) is 6, 9, 8, 8, 8 and 8 respectively. The discretized factors were used as input into the GDM, and the spatial consistency between each comprehensive factor and GSN slope was tested.

*2.3.4 Geographically weighted regression model (GWR)*

Our study uses the Geographical Weight Regression model (GWR) (Brunsdon et al., 1996) to explore spatial correlations between comprehensive factors and GSN slope (local R square), as well as the sensitivity of GSN slopes to different drivers (expressed by the regression slope). Assuming that the drivers of GSN slopes are non-stationary and vary across the study area, a global model has often limited explaining power to express the relationship between GSN slopes and the comprehensive factors. This is particularly true for the Chinese karst area, where different climate conditions and geology/project regions (Fig. 1) provide different settings (expressed by project regions). Here GWR can assess the local relationships between the non-stationary comprehensive factors and GSN slopes by allowing the value of local parameters ($\beta_k(\mu,\nu)$) to change with the geographical location ($x_k$), as shown in Eq. (3) (Peng et al. 2017, Zhang et al. 2017b):

$$y_i = \beta_{i0}(\mu_i,\nu_i) + \sum_{k=1} \beta_{ik}(\mu_i,\nu_i)x_{ik} + \varepsilon_i$$

(3)

Where $(\mu_i,\nu_i)(i=1,2,3,\cdots,n)$ is the spatial location of the sample $i$; $y_i$ represents the dependent variable (it refers to GSN slope in this study); $x_{ik}$ represents the independent variables, which refers to the 6 comprehensive factors (PCs) here; $\beta_{i0}(\mu_i,\nu_i)$ and $\beta_{ik}(\mu_i,\nu_i)$ represent the constant estimates and parametric estimates of the sample $i$, respectively; $\varepsilon_i$ is the random error.

As multicollinearity between environmental factors has been eliminated by the PCA, we built multiple regression models using ordinary least squares (OLS) and GWR by taking the six PCs as independent and GSN slope as dependent variable. The statistical results of the OLS model are compared with those of GWR in Table S4. We use the Akaike Information Criterion (AICc), Cross Validation (CV) and Bandwidth Parameter (PAR) to determine the extent of the kernel (Table S4), showing that the GWR model is superior to OLS.

## 3. Results

*3.1 Temporal and spatial variation of vegetation trends*

Over 1982-2016, southwest China karst has seen a general greening trend, which is stronger during the conservation period (2001-2016), as compared to the reference period (1982-2000) (Table S1). In the karst area, trends changed from negative to positive between the periods, which was not the case for non-karst areas (Fig. 3a, b). Most greening areas are found in Guangxi and Guizhou, while vegetation cover in Yunnan shows a decreasing trend in the conservation period (Fig. 3a). Spatial differences of GSN slopes also exists in different karst landforms during the conservation period. The GSN slope in the Karst Peak-Cluster Depression (II) shows the largest share of significant increase (9.43%), followed by the Karst Peak Forest Plain (I) (3.88%), Karst Plateau (VI) (3.39%), Karst Gorge (V) (3.38%) and Karst Trough Valley (III) (3.11%). Only a small percentage of the pixels are found to have a significant decreasing trend in karst areas (less than 5%). The GSN slope in the conservation period is overall positive in most parts of the karst region except the Middle-High Hill region (IV) and the Karst Fault Basin (VII) with only 3.60% and 13.91% showing a significant increase (Fig. 3c, d). More than half part of the areas in the Karst Trough Valley (III) (50.32%) was increased significantly, and the Karst Plateau (VI) (48.19%), Karst Peak Forest Plain (I) (46.48%) and Karst Peak-Cluster Depression (II) (45.13%) were dominated by significantly increase.

GSN slopes show significant ($p<0.05$) spatial autocorrelation for both periods but with a much stronger spatial clustering during the conservation period. The global Moran's I and Z(I) values before the implementation of eco-engineering are 0.60 and 17.73, respectively, while global Moran's I and Z(I) during the conservation period are 0.85 and 25.04, respectively. Furthermore, the increase in GSN slope is spatially not homogeneous. Areas with strongly positive GSN slope clusters are found in non-karst areas in Yunnan and Karst Peak Forest Plain during the reference period and Karst Trough Valley, Plateau, Gorge, Karst Peak Forest Plain and Peak-Cluster Depression during the conservation period. Decreasing slopes are mainly concentrated in the Karst Fault Basin and Middle-High Mountains (Fig. 3e.f). The following sections will further analyze different natural and anthropogenic factors causing these spatial variations in GSN slopes.

## 3.2 Climatic conditions and trends in the study area

Over 2001-2015, mean annual temperature increased significantly ($p<0.05$) and mean annual precipitation decreased significantly ($p<0.05$) in the western part of the study area (Fig. 4c, d), which is consistent with the spatial distribution of vegetation decrease shown in Fig. 3d. Areas with no significant changes in climate variables are dominated by vegetation increase. There are three climate types with significant hydrothermal differences in the study area. The Middle-High Mountains and Karst Fault Basin belong to the plateau climate and Southwest monsoon climate zone, respectively, with a relatively arid and cold climate, both the mean annual precipitation (MAP) and mean annual temperature (MAT) are significantly (p<0.05) lower than in other karst areas. GSN slopes in these two areas are significantly lower than in other karst areas belonging to the East Asia monsoon climate zone with relatively abundant hydrothermal resources, such as in the Karst Peak Forest Plain, Karst Peak-Cluster Depression, Karst Gorge, Karst Trough Valley and Karst Plateau. However, in spite of annual precipitation and temperature vary greatly over the karst area belonging to East Asia monsoon climate zone (Fig. 4a, b), GSN slopes are relatively similar between these project regions (Fig. 3d). Consequently, climate can have a significant impact on the vegetation trends but are not the main reasons for vegetation change in karst areas with relatively favorable climate conditions.

*3.3 Impact of environmental factors on vegetation trends*

Using GDM, we estimated the magnitude of the impact of the six comprehensive factors on GSN slopes at the 95% confidence level. The PD values show the importance of the comprehensive factors in different project regions (Table 2). The influences of human and natural factors on GSN slopes vary greatly in different project regions: Values reflecting human activities are higher in the Karst Peak Forest Plain (0.80), Karst Middle-High Mountains (0.79) and Karst Gorge (0.74). Similarly, PD values reflecting socio-economical conditions are highest in the Karst Plateau (0.70) and lowest in the Karst Peak-Cluster Depression (0.06). GSN slopes in Middle-High Mountains were greatly influenced by human factors: eco-engineering (0.81) > interference of human activities (0.79) > socio-economic (0.76). The order of the influence of each factor on GSN slopes in the Karst Peak-Cluster Depression is: soil texture (0.34) > soil nutrients (0.33) > human activities (0.25) > ecological engineering (0.18) > socio-economic (0.06). Climate conditions have little influence on vegetation trends in karst areas (most PD values are less than 0.10), but are the dominant factor affecting vegetation trends in non-karst areas (PD=0.68). Eco-engineering plays an important role for vegetation trends in the Karst Middle-High Mountains (0.81), Karst Peak Forest Plain (0.42), Karst Gorge (0.29) and Karst Peak-Cluster Depression (0.18), and it is the dominant factor in the Karst Trough Valley (0.75), Karst Fault Basin (0.54) and Karst Plateau (0.32), but not in the non-karst areas (PD=0.05).

The contrasting influences of climate condition and eco-engineering on vegetation trends in karst and non-karst areas was further confirmed by GWR models. The strength of correlations between comprehensive factors and GSN slopes varies through space, expressed by the range of values shown as boxplots in Fig. 5. The influence of eco-engineering on GSN slopes in karst areas is significantly ($p<0.05$) greater than that of climate conditions, and the average local $R^2$ values are 0.16 and 0.09, respectively. However, the opposite is the case in non-karst areas, where the average local $R^2$ for eco-engineering is 0.11, and for climate condition it is 0.17. High correlations (local $R^2>0.5$) between eco-engineering and GSN slopes are found in the Karst Trough Valley, while correlations with climate condition are higher in non-karst areas, both is consistent with the GDM based analyses (Fig. 4).

Both the regression slopes of the six PCs and the strength of the correlations (local $R^2$) are spatially heterogeneous, in other words, the sensitivity of GSN slopes to the comprehensive factors is not spatially homogeneous. The regression slopes of the natural factors expressed by climate condition (PC1) and soil nutrients (PC5) show greater variations than the anthropogenic factors (PC2,4,6), both for karst and non-karst areas (Fig. 5d, e, f). In addition, GSN increases together with eco-engineering. All regression slopes related to eco-engineering are positive while half of the slopes of climate condition are negative in the karst areas (Fig.5e). A total of 85.39% of the regression slopes related to climate condition are positive in non-karst areas (Fig.5f), which confirms that decreases in rainfall coincide with negative vegetation trends (Fig.3d, Fig.4c, d).

We then studied the interaction between the comprehensive factors to test how the effects of eco-engineering are enhanced by other factors. The 5 interacting factor pairs that supported eco-engineering to increase GSN are listed in Table 2. Enhancements were detected in all project regions and for all factors. The impact of eco-engineering on vegetation trends was nonlinearly enhanced in Karst Peak-Cluster Depression and Karst Plateau, while a double-synergy effect (bi-enhanced) occurred in Karst Middle-High Mountains. Enhancements vary greatly between the different landforms and it is clear from Table 2 that decision-makers should pay attention on the natural and social conditions in different project regions to determine the measures of eco-engineering management.

## 4. Discussion

*4.1 Dominant factors in different karst project regions*

We studied long-term changes and spatial patterns in vegetation cover during 1982 to 2016 and confirm the greening trend in the karst region of south China (Zhu et al. 2016, Tong et al. 2017, Brandt et al. 2018, Tong et al. 2018). After the implementation of ecological engineering projects starting in the year 2000, vegetation growth accelerated, which is in line with previous studies (Tong et al. 2016, Brandt et al. 2018, Tong et al. 2018, Tong et al. 2020). Furthermore, we show that areas of vegetation increase cluster mainly in the karst region of Guangxi and Guizhou (Tong et al. 2016), while a decline in vegetation growth was observed in the Karst Middle-High Mountains, Karst Fault Basin and the non-karst areas of Yunnan.

Many studies have suggested that the spatio-temporal patterns of regional vegetation trends are the consequence of climate change and land use conversions (Piao et al. 2015, Qian et al. 2019, Zhuge et al. 2019). Here we use six comprehensive factors reflecting climate conditions and human activities and give statistical evidence on the importance of the factors on vegetation trends. We show that climate conditions and a decrease in rainfall and increase in temperature align with the negative vegetation trends in the non-karst areas of Yunnan (Brandt et al. 2018, Wang et al. 2020, Zhao et al. 2020a). However, in most karst areas, such as the Peak Forest Plain, Peak-Cluster Depression, Karst Gorge, Karst Trough Valley and the Karst Plateau, there was no significant change in rainfall and temperature (Fig.4c,d), and climate conditions were found to be irrelevant for the patterns of recent vegetation trends. We further showed that most of the karst area is located in the East Asia monsoon climate zone, having a stable and favorable climate over the study period, which benefits eco-engineering. We further provide statistical evidence that eco-engineering is indeed the dominant factor for vegetation increase in fragile karst areas, such as the Karst Fault Basin and the Middle-High Mountains, where hydrothermal conditions are rough and the environment is sensitive to human activities (Tao et al. 2020).

Human activities determine the vegetation trends in the Karst Peak Forest Plain and Karst Gorge. Paddy soils in the Karst Peak Forest Plain cover 12%, which is the highest percentage compared to other project regions. In spite of the high degree of land use, high effectiveness of eco-engineering was found in the Karst Peak Forest Plain (Tong et al. 2017). Population concentrates in flat terrain and desertification in the Karst Peak Forest Plain mainly occurs on sloping hills, making these areas a hot-spot for recovery measures. The degree of land use was lower in the Karst Gorge, but high altitude and steep slopes make the landscape vulnerable and difficult to recover (Zhao et al. 2020a), requiring reasonable measures.

Soil properties dominate vegetation trends in the Karst Plateau and Peak-Cluster Depression, which are areas where severe rocky desertification is reported (Jiang et al., 2014, Zhao et al., 2020b). Karst Plateaus have a high coverage with paddy soils (10%), making farmers independent of sloping lands and supporting vegetation growth. However, Karst Peak-Cluster Depressions have little arable land (paddy soils cover only 5%) and most farmland is located on sloping hills, making rocky desertification a serious issue in these areas (Wang 2016). Zhang and Wang (2009) showed that a shortage of mineral nutrients in shallow soil causes a low vegetation productivity in karst mountain areas. Here, vegetation restoration assisted by mineral fertilizers may be more effective in soil fissures (Zhang and Wang 2009, Peng et al. 2020). Furthermore, the calcareous soils are fertile and support vegetation growth if human interference is kept low, making to reduction of farmlands on sloping land an effective tool to restore vegetation (Gao and Wang 2019).

Ecological engineering was the dominant factor in Karst Trough Valleys and Fault Basins. In spite of a moderate project effectiveness detected by Tong et al. (2017), we found ecological conservation causing vegetation increases in these areas, consistent with Xu et al. (2019). Soil degradation in the Karst Fault Basin hinders vegetation growth if human interference is high (Xu et al. 2019, Shen et al. 2020). Increasing vegetation cover, reduced human disturbance and measures to increase soil nutrients may help soils to recover (Shen et al. 2020). The ecological environment of Karst Middle-High Mountains is fragile and difficult to recover (Xiong et al. 2014). While the transformation of farmland into grassland and forests helps to accumulate soil organic matter (Liu and Huang 2005, Tong et al. 2017), the establishing of natural reserves and mountain closures (Zhang et al. 2017) would be most effective in Karst

Middle-High Mountains, where population is low (Liu and Huang 2005, Jiang et al. 2014, Xiong et al. 2014, Zhao et al. 2020a, Zhao et al. 2020b).

## 4.2 Interacting effects on vegetation trend in karst areas

The karst area is characterized by steep slopes, shallow soils, dense population and severe poverty (Wang et al. 2008, Zhang et al. 2017, Hu et al. 2018). Therefore, afforestation should consider the local conditions and ecological and economic benefits of the areas (Cao 2011). The implementation of eco-engineering should focus on areas with limited land resources, as the economic and ecological value of these areas could be considerably increased by appropriate measures (Wang et al. 2017b, Qiu et al. 2020). A variety of different vegetation trends between project regions results from different interactions between natural factors and human activities. Quantifying these interactions is helpful for policymakers and stakeholders to adopt ecological restoration measures to local settings (Shinn 2016, Zhang 2020). For example, because of unfavorable natural conditions, GSN slope has shown a decrease trend in Middle-High Mountains and the Karst Fault Basin. Cold temperatures limit vegetation growth in the protected areas along the rivers in Yunnan, and a decrease in precipitation causes a negative trend in vegetation cover. Strengthening land use management and reducing human disturbance are effective measures to prevent ecological degradation in Middle-High Mountains (Tong et al., 2020), which illustrates the synergic influence of eco-engineering and natural factors on vegetation trends. Affected by the dry-hot valley climate and global warming, unpredictable rainfall and extreme solar radiation results in water-shortage in the Karst Fault Basin and Gorge, leading to patches of vegetation surrounded by bare soil (Yuan et al. 2020). However, human factors such as urbanization and large-scale hydropower development along the Jinsha River (Wu et al.

2020b) explain the non-linearly enhanced influence of eco-engineering and human activities on vegetation trends. In addition, the restoration of vegetation in areas with more cultivated land resources (like the Karst Peak Forest Plain) should focus on supporting the growth of new and precious plant species like the slow-growing Dalbergia odrifera instead of planting fast growing forests (Jiang et al., 2014, Wang et al., 2016, Zhang et al., 2020). In areas of high human pressure (like the Karst Plateau, Trough Valley and most parts of the Peak-Cluster Depression) natural regeneration on slopes should be combined with plantations on mountain foots (Wang 2016, Qiu et al. 2020).

## *4.3 Improvement and uncertainty of this study*

Many studies have studied the drivers of vegetation change in China karst (Tong et al. 2017, Tao et al. 2018, Li et al. 2019, Xu et al. 2019, Zhang et al. 2019, Hou and Gao 2020, Liu et al. 2020, Qiu et al. 2020, Wu et al. 2020a, Zhang et al. 2020, Zhao et al. 2020a). Our study contributes to the field using comprehensive factors without multicollinearity (using PCA) and simplified models which are easy to replicate. Moreover, we applied both GDM and GWR models which work differently but lead to similar results, showing that climate conditions dominant vegetation trends in non-karst areas, while it was eco-engineering in most karst areas. While both models consider spatial heterogeneity, GDM includes the interacting effects between environmental factors on vegetation trends, and the GWR model quantifies the sensitivity of vegetation trends to the factors.

We identified drivers affecting vegetation trends at a regional scale, highlighting the role of large scale ecological conservation projects. There are still several aspects that can be improved in future research. First, the scale and zoning (or aggregation) problems are part of spatial analysis (Ju et al. 2016) and can lead to different results due to different aggregation sizes, methods or spatial arrangements (Jelinski and Wu 1996). For the scale, ecological engineering data are only available in administrative units (Tong et al. 2017), which limits all analyses to the scale of counties. For the zoning effect, using the most appropriate discretization method to stratify continuous variables into several categories is challenging, because these methods do not have standardized rules (Cao et al. 2013, Zhao et al. 2017). Second, there is a loss of information when performing PCA for dimension reduction, and the quality of the data sources representing the comprehensive factors varies greatly.

## 5 Conclusion

This study analyzed temporal vegetation trends and their driving forces in the Guangxi-Yunnan-Guizhou region in southern China. The results show that vegetation growth in the karst areas increased strongly after the implementation of conservation projects after the year 2000, which was not the case for non-karst areas. Climate conditions and a decrease in rainfall were responsible for low vegetation growth in non-karst areas, while eco-engineering was the main cause for vegetation increase in karst areas. Our study gives quantitative and statistical evidence that large scale ecological conservation projects are the main reason for the greening trend in south China karst, while climate is the dominant driver of negative vegetation trends in non-karst areas. Furthermore, interactions between natural factors and human activities enhance vegetation trends in all project regions, indicating the importance of considering climate and geological settings when planning and evaluating eco-engineering measures.

## Acknowledgements

This study was funded by the Strategic Priority Research Program of Chinese Academy of Sciences (grant number XDA19050502); the National Natural Science Foundation of China (grant number 41930652, U20A2048); the National Key Research and Development Program of China (grant numbers 2016YFC0502400, 2016YFC0502501, 2018YFD1100103); the CAS Key Laboratory of Agro-ecological Processes in Subtropical Region (grant numbers ISA2018202); Marie Curie fellowship (grant number 795970) and the National Natural Science Foundation of China (grant numbers 41501206). We thank GIMMS NDVI Group for producing and sharing the NDVI dataset. Resource and Environment Data Cloud Platform, Institute of Geographical Sciences and Natural Resources Research, CAS is thanked for producing and sharing the climatic, soil and socio-economic data. We thank the Forest Bureau of the Yunnan, Guizhou and Guangxi Provinces for providing the statistical data.

### Table 1. List of datasets used in this study

| Attribute | Environmental Indexes | Description and resolution | Provider |
|---|---|---|---|
| Climate factors (Zhang et al. 2017, Lv et al. 2019) | ＞0℃ accumulated temperature | meteorological data (1km) | http://www.resdc.cn |
| | ＞10℃ accumulated temperature | meteorological data (1km) | |
| | Aridity | meteorological data (1km) | |
| | Annual mean temperature | annual mean temperature and precipitation from | |
| | Annual mean precipitation | 2001 to 2015 (1km) | |
| Soil properties (Zhang et al. 2017) | Clay content | soil texture (1km) | http://www.resdc.cn |
| | Silt content | soil texture (1km) | |
| | Sand content | soil texture (1km) | |
| | Soil nitrogen storage | nitrogen storage (5km) | http://westdc.westgis.ac.cn |
| | SOC on soil type | 1:1M scale China soil database | http://sourcedb.issas.cas.cn/ |
| | SOC on vegetation type | 1:1M scale China soil database | http://sourcedb.issas.cas.cn/ |
| Geological background (Tong et al. 2016) | DEM | digital elevation model (30 m) | http://www.radi.cas.cn/ |
| | Slope | Derived from DEM in ArcGIS 10.2 (30 m) | |
| Hydrologic condition (Lv et al. 2019) | River density | 1:100M national fundamental geographic map | China Geological Survey |
| Ecological engineering (Tong et al. 2017, Han et al. 2019) | Project area | Statistical data at county scale | Forestry Bureau of the Yunnan, Guizhou, and Guangxi Provinces |
| | Capital input | Statistical data at county scale | |
| Human activities (Tong et al. 2016, Han et al. 2019, Lv et al. 2019) | Population density | population spatial distribution (1km) | http://www.resdc.cn |
| | Gross Domestic Product | gross domestic product (GDP) (1km) | |
| | Road density | Road length per unit area at county scale | |
| | Land use types | land use map in 2005, 2010, 2015 (30 m) | http://www.radi.cas.cn/ |

**Table 2 PD values in different project regions**

| | PC4 | PC1 | PC2 | PC3 | PC5 | PC6 | PC4∩PC1 | PC4∩PC2 | PC4∩PC3 | PC4∩PC5 | PC4∩PC6 |
|---|---|---|---|---|---|---|---|---|---|---|---|
| **I** | 0.42 | 0.08 | **0.80** | 0.25 | 0.46 | 0.20 | 0.50 ↑ | 0.92 ↑ | 0.98 ↑↑ | 0.70 ↑ | 0.72 ↑↑ |
| **II** | 0.18 | 0.09 | 0.25 | **0.34** | 0.33 | 0.06 | 0.36 ↑↑ | 0.75 ↑↑ | 0.70 ↑↑ | 0.74 ↑↑ | 0.68 ↑↑ |
| **III** | **0.75** | 0.04 | 0.44 | 0.30 | 0.07 | 0.35 | 0.82 ↑↑ | 0.89 ↑ | 0.84 ↑ | 0.86 ↑↑ | 0.94 ↑ |
| **IV** | 0.81 | 0.17 | 0.79 | 0.18 | **0.82** | 0.70 | 0.94 ↑ | 1.00 ↑ | 1.00 ↑ | 0.96 ↑ | 1.00 ↑ |
| **V** | 0.29 | 0.34 | **0.74** | 0.61 | 0.47 | 0.26 | 0.85 ↑↑ | 0.96 ↑ | 0.94 ↑↑ | 0.80 ↑↑ | 0.74 ↑↑ |
| **VI** | **0.32** | 0.15 | 0.07 | 0.12 | 0.28 | 0.21 | 0.60 ↑↑ | 0.57 ↑↑ | 0.80 ↑↑ | 0.69 ↑↑ | 0.71 ↑↑ |
| **VII** | **0.54** | 0.08 | 0.07 | 0.25 | 0.38 | 0.23 | 0.60 ↑ | 0.67 ↑↑ | 0.77 ↑ | 0.75 ↑ | 0.90 ↑↑ |
| **Non-karst** | 0.05 | **0.68** | 0.45 | 0.10 | 0.20 | 0.30 | 0.79 ↑↑ | 0.59 ↑↑ | 0.45 ↑↑ | 0.61 ↑↑ | 0.53 ↑↑ |

Note: ↑↑ means non-linear enhancement; ↑ means bi-enhancement. I: Peak Forest Plain; II: Peak-Cluster Depression; III: Karst Trough Valley; IV: Middle-High Mountains; V: Karst Gorge; VI: Karst Plateau; VII: Karst Fault Basin. PC1: Climate condition. PC2: Human activities. PC3: Soil texture. PC4: Eco-engineering. PC5: Soil nutrients. PC6: Socio-economic.

**Figure captions**

**Fig.1. Location and classification of the different project regions.** I: Peak Forest Plain; II: Peak-Cluster Depression; III: Karst Trough Valley; IV: Middle-High Mountains; V: Karst Gorge; VI: Karst Plateau; VII: Karst Fault Basin.

**Fig.2. Flowchart showing the data and methods used in this study.** GWR: Geographical weighted regression model; GDM: Geographical detector model; GSN: Growing Season NDVI.

**Fig.3. Spatial distributions of GSN slopes. a. GSN slopes for 1982-2000. b. Same for 2001-2016. c. The area ratio (percentage cover) of GSN slopes during the conservation period (2001-2016) in different project regions. d. Boxplots showing the distribution of GSN slopes over the different project regions. e. Spatial aggregation (via autocorrelation analysis) of positive and negative trends in GSN slope for 1982-2000. f. Same for 2001-2016.** The project regions are Peak Forest Plain (I), Peak-Cluster Depression (II), Karst Trough Valley (III), Middle-High Mountains (IV), Karst Gorge (V), Karst Plateau (VI), Karst Fault Basin (VII) and Non-karst area (the shadow area).

**Fig.4. Climate conditions and their trends for 2001-2015 in different project regions. a. Mean annual precipitation (MAP), b. Mean annual temperature (MAT). c. Significant ($p<0.05$) slopes of mean annual temperature for 2001-2015. d. Significant ($p<0.05$) slopes of mean annual precipitation for 2001-2015.** The project regions are Peak Forest Plain (I), Peak-Cluster Depression (II), Karst Trough Valley (III), Middle-High Mountains (IV), Karst Gorge (V), Karst Plateau (VI), Karst Fault Basin (VII) and Non-karst area (the shadow area).

**Fig.5. Local R squares and slopes of GWR in different regions.** PC1: Climate condition. PC2: Soil texture. PC3: Eco-engineering. PC4: Human activities. PC5: Soil nutrients. PC6: Socio-

economic.

Figure1  Click here to access/download;Figure;fig1_revise.tif

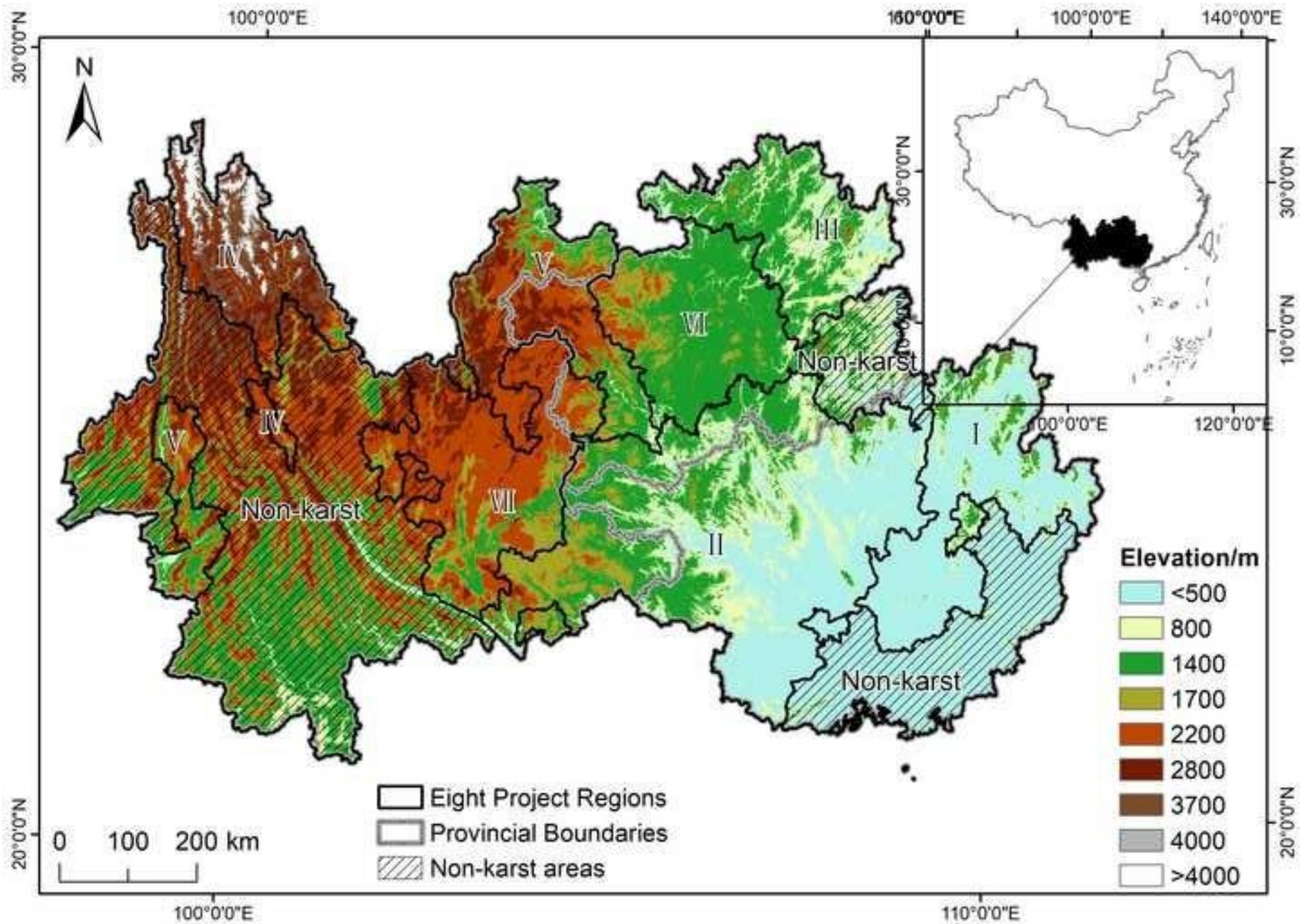

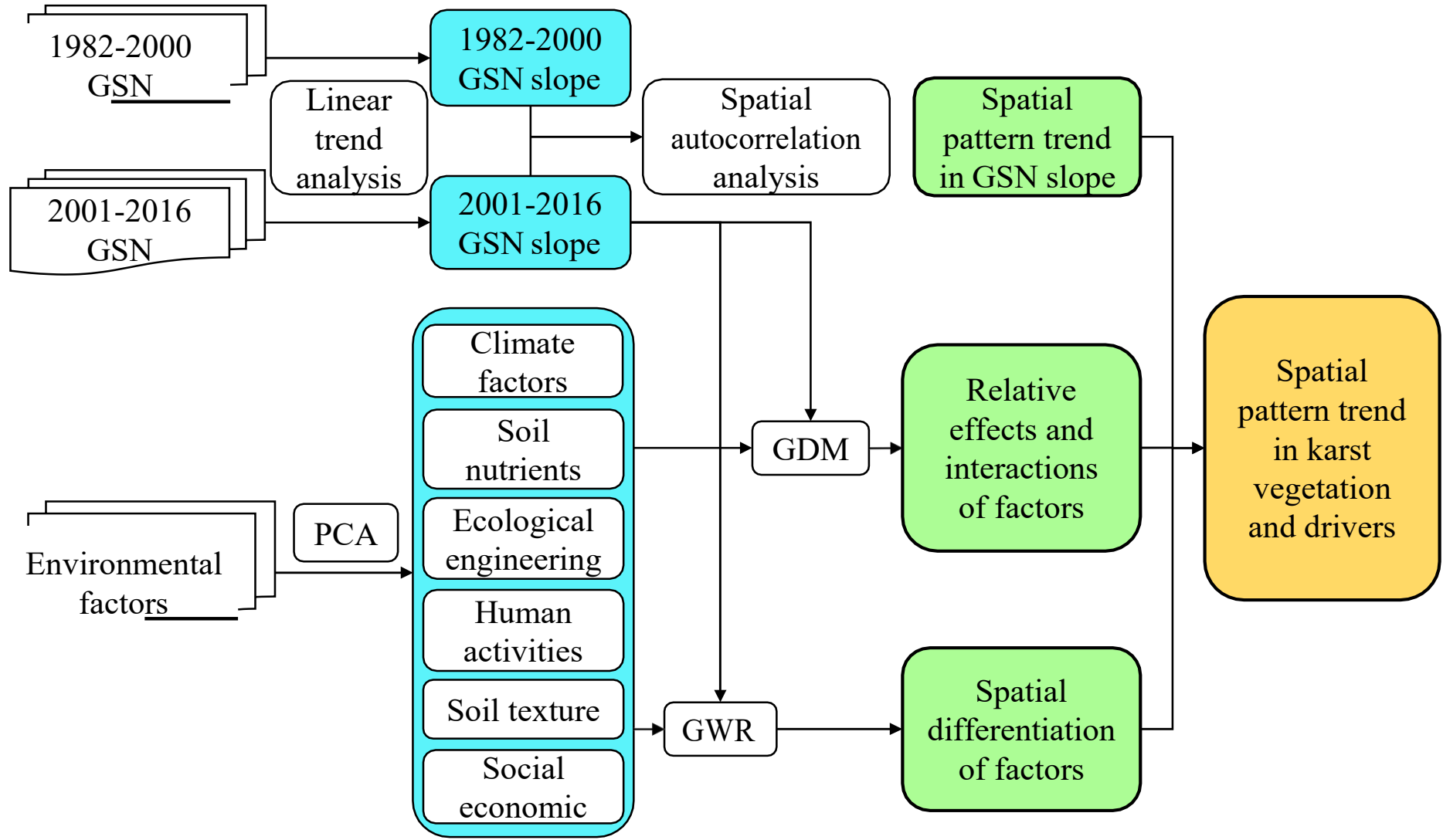

Figure3a

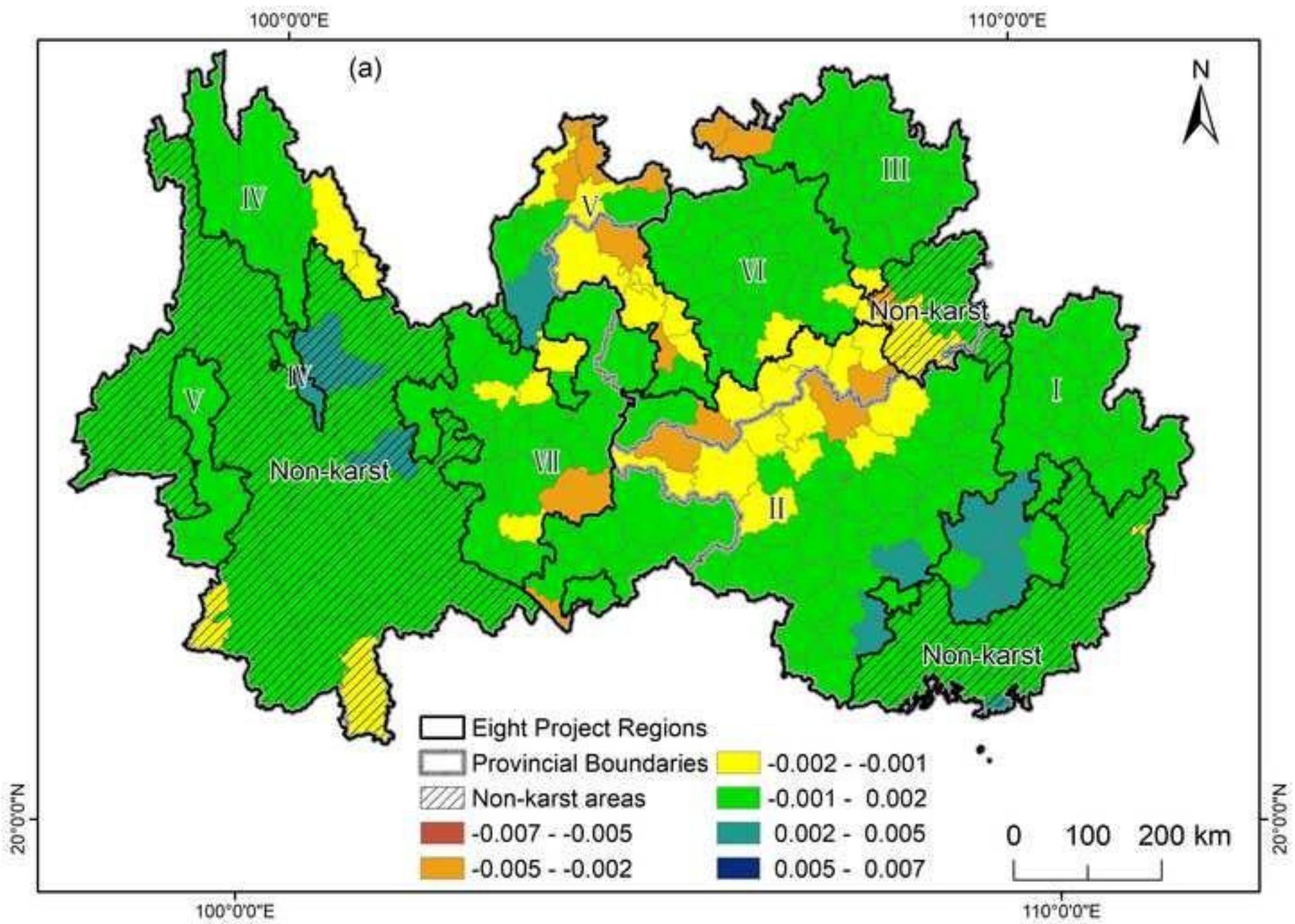

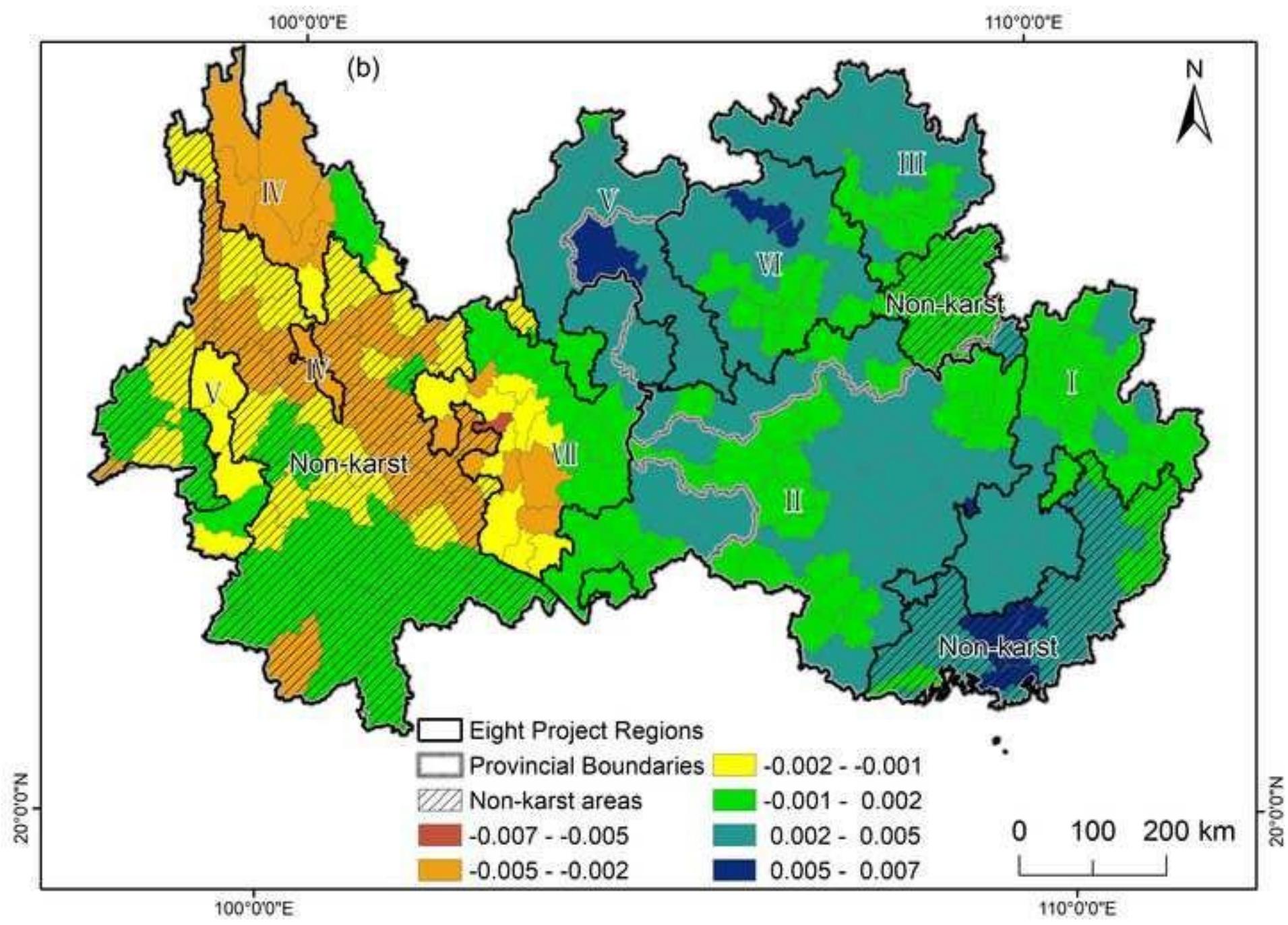

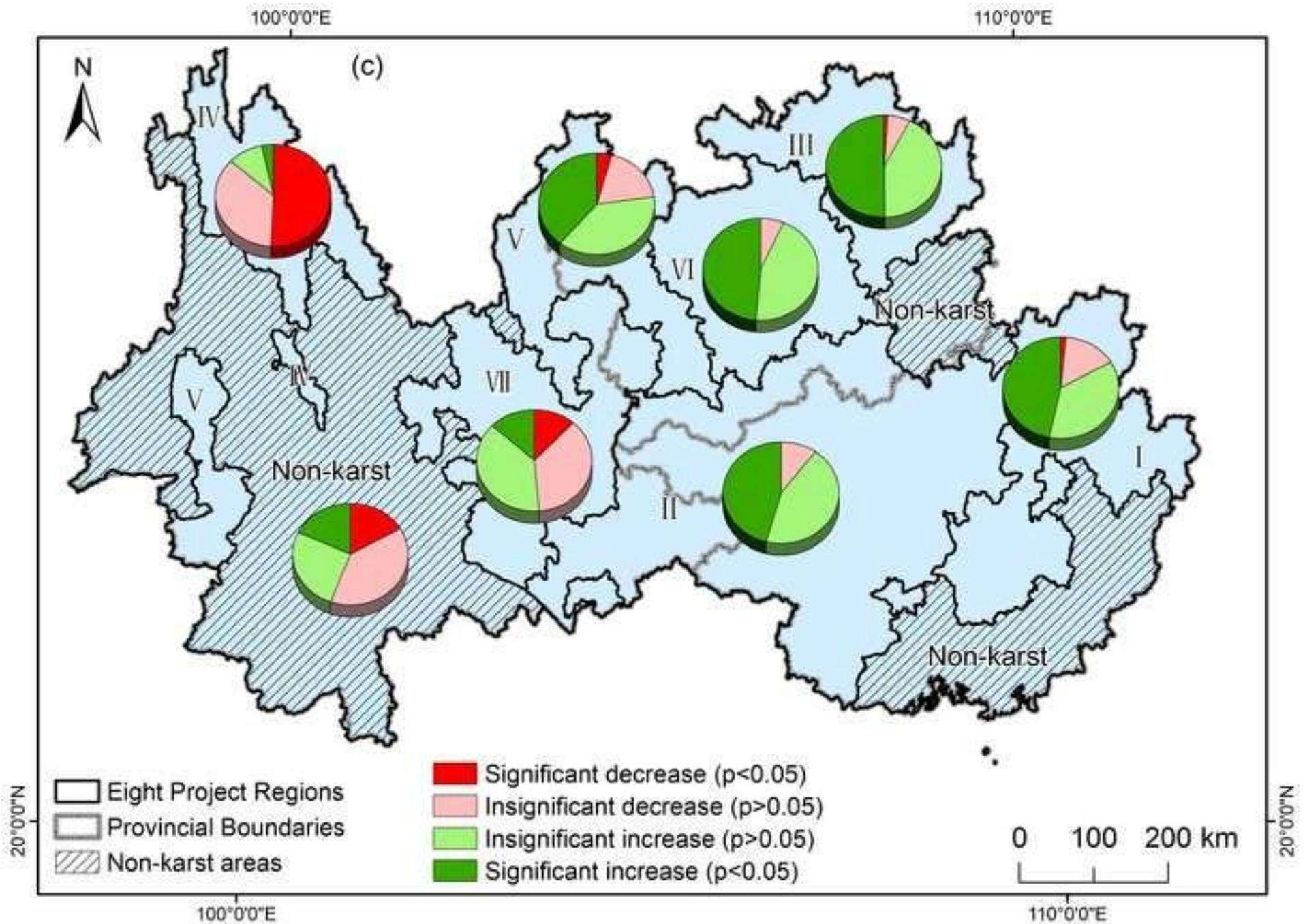

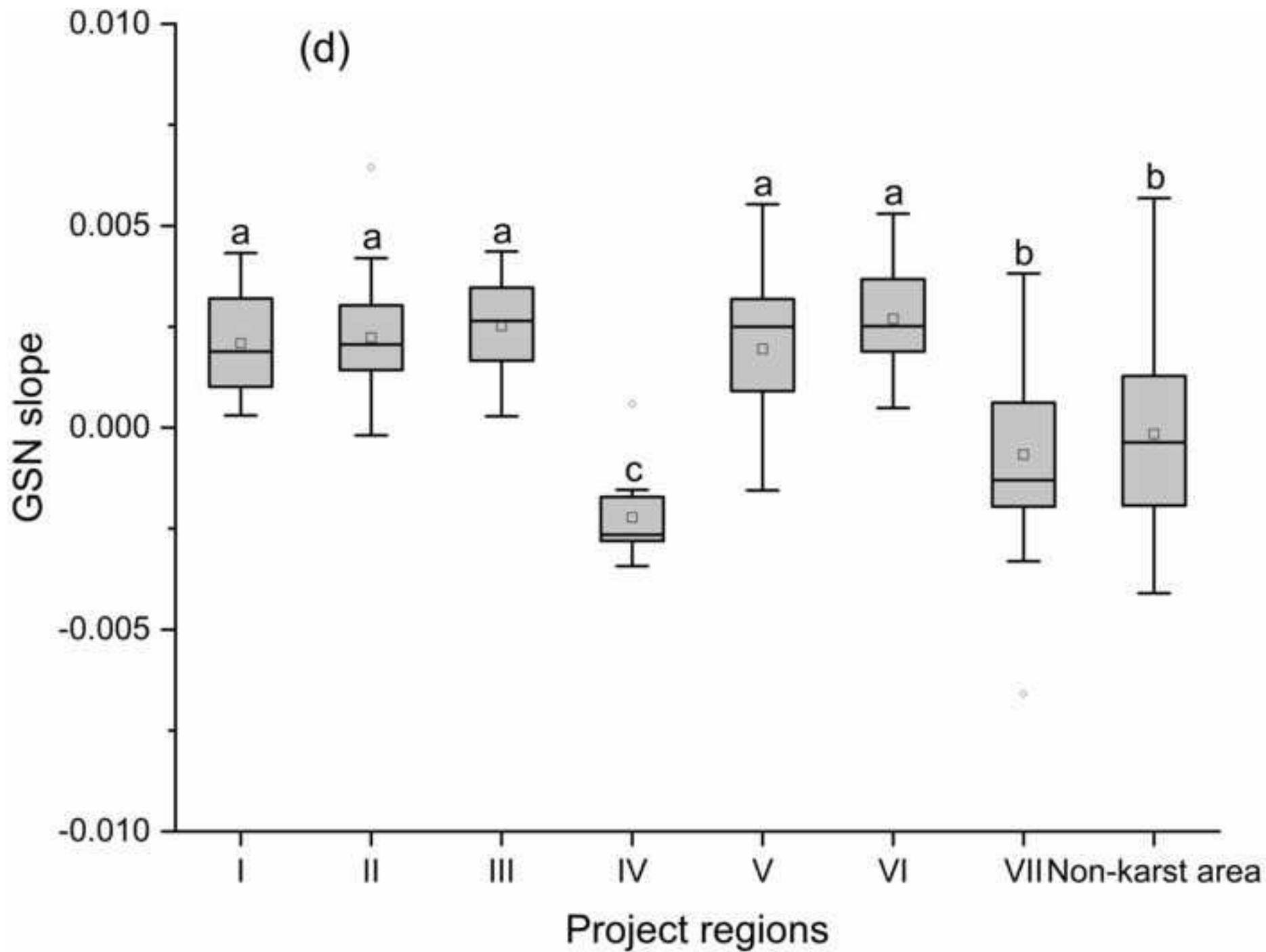

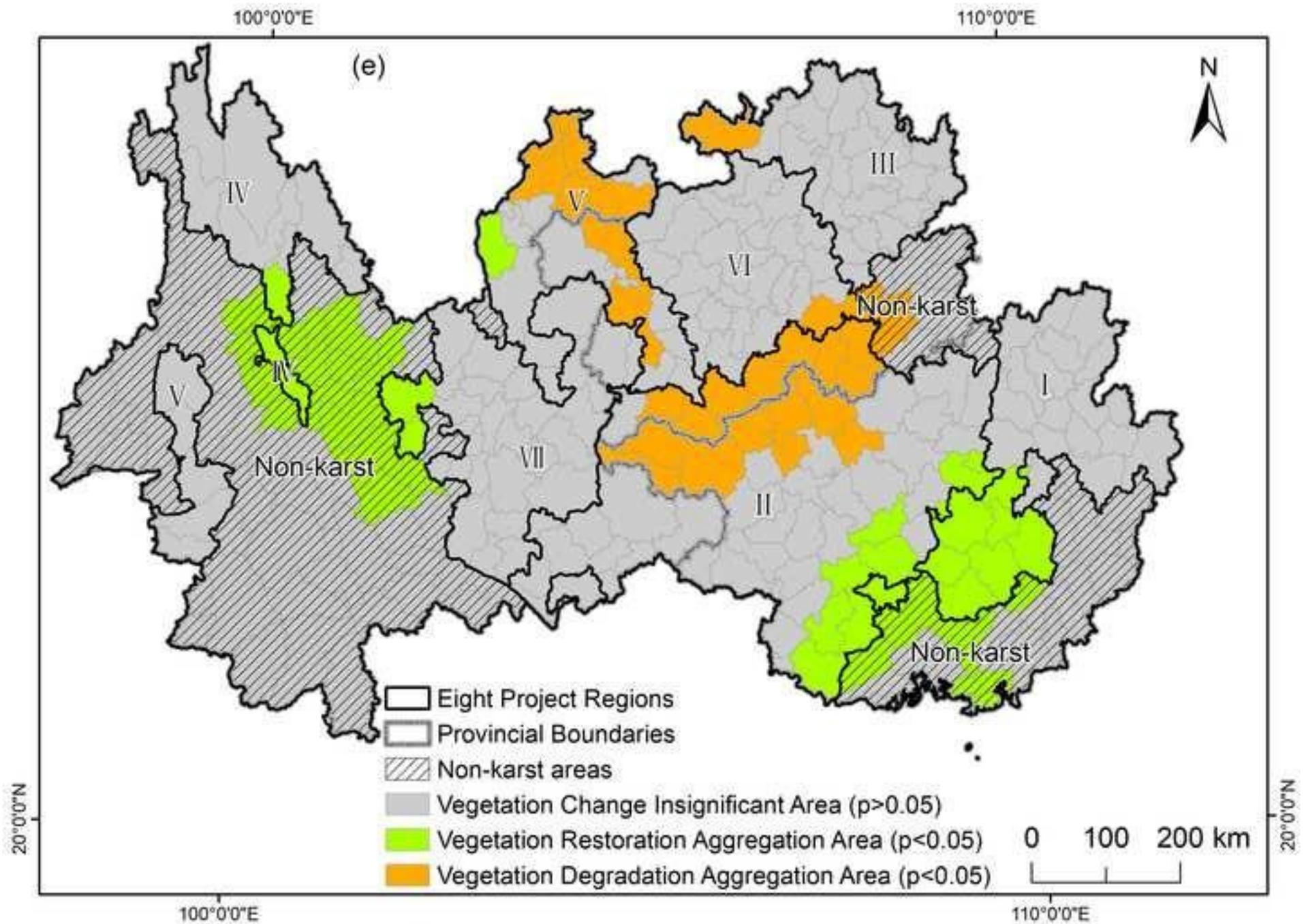

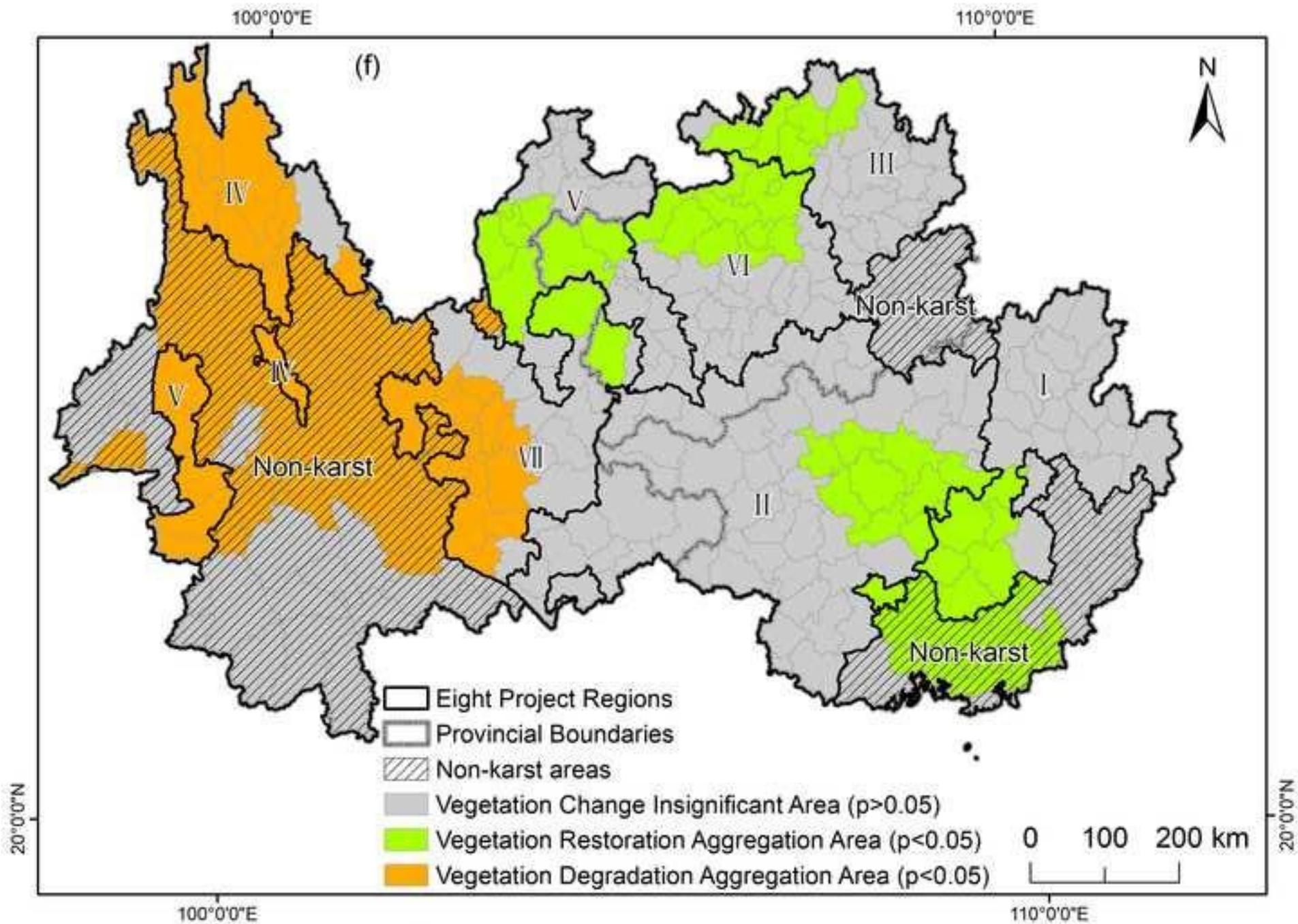

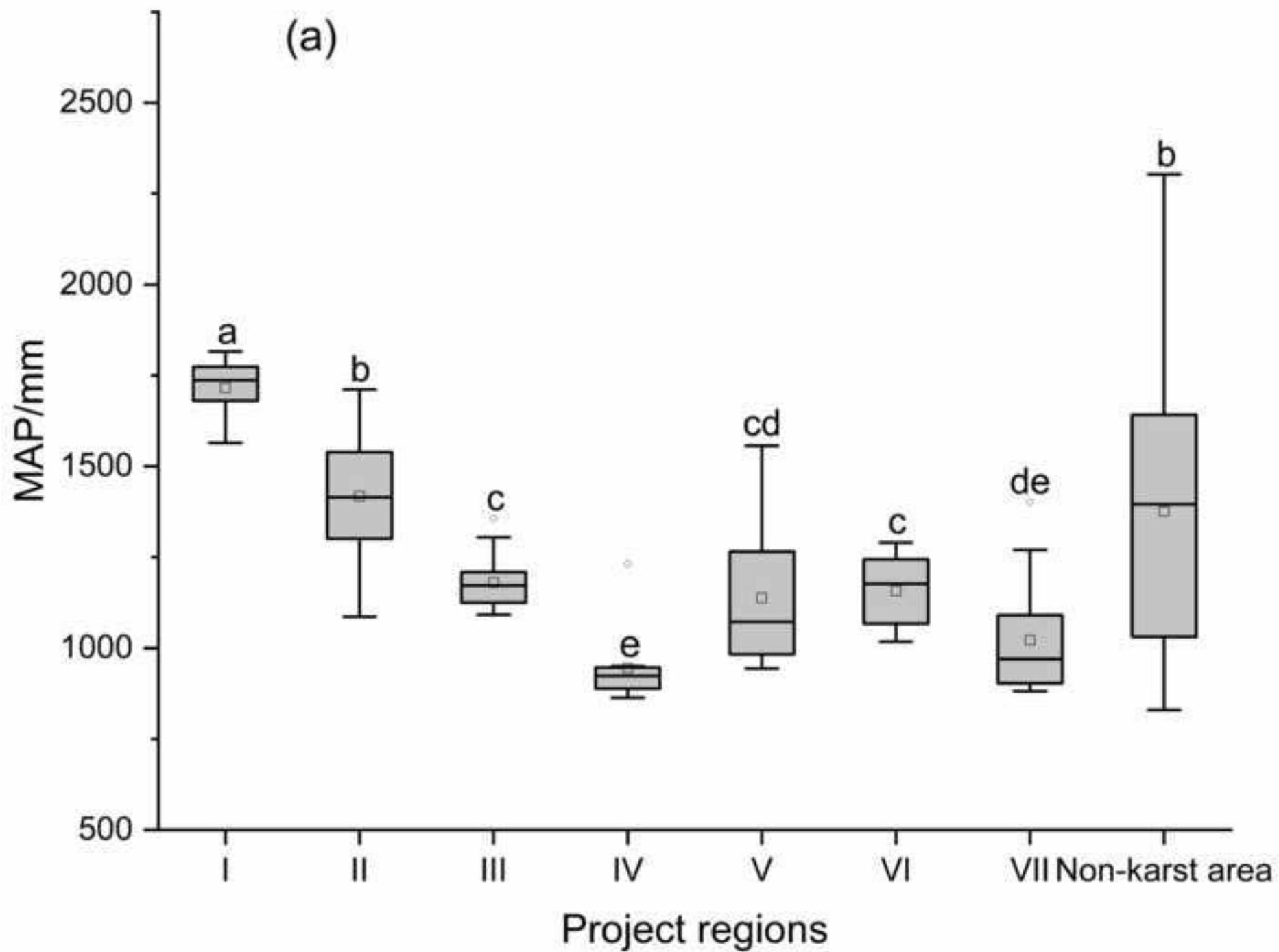

 

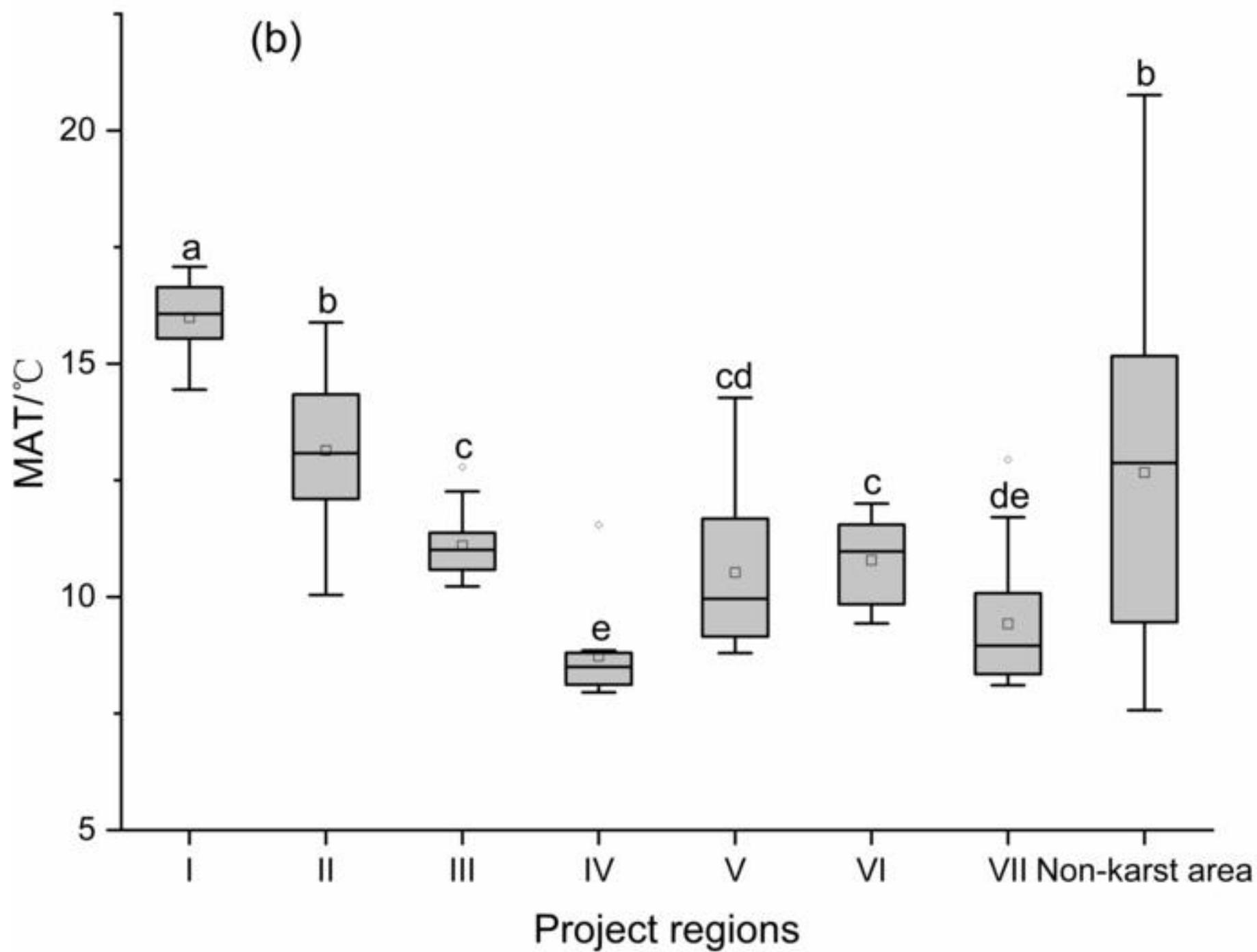

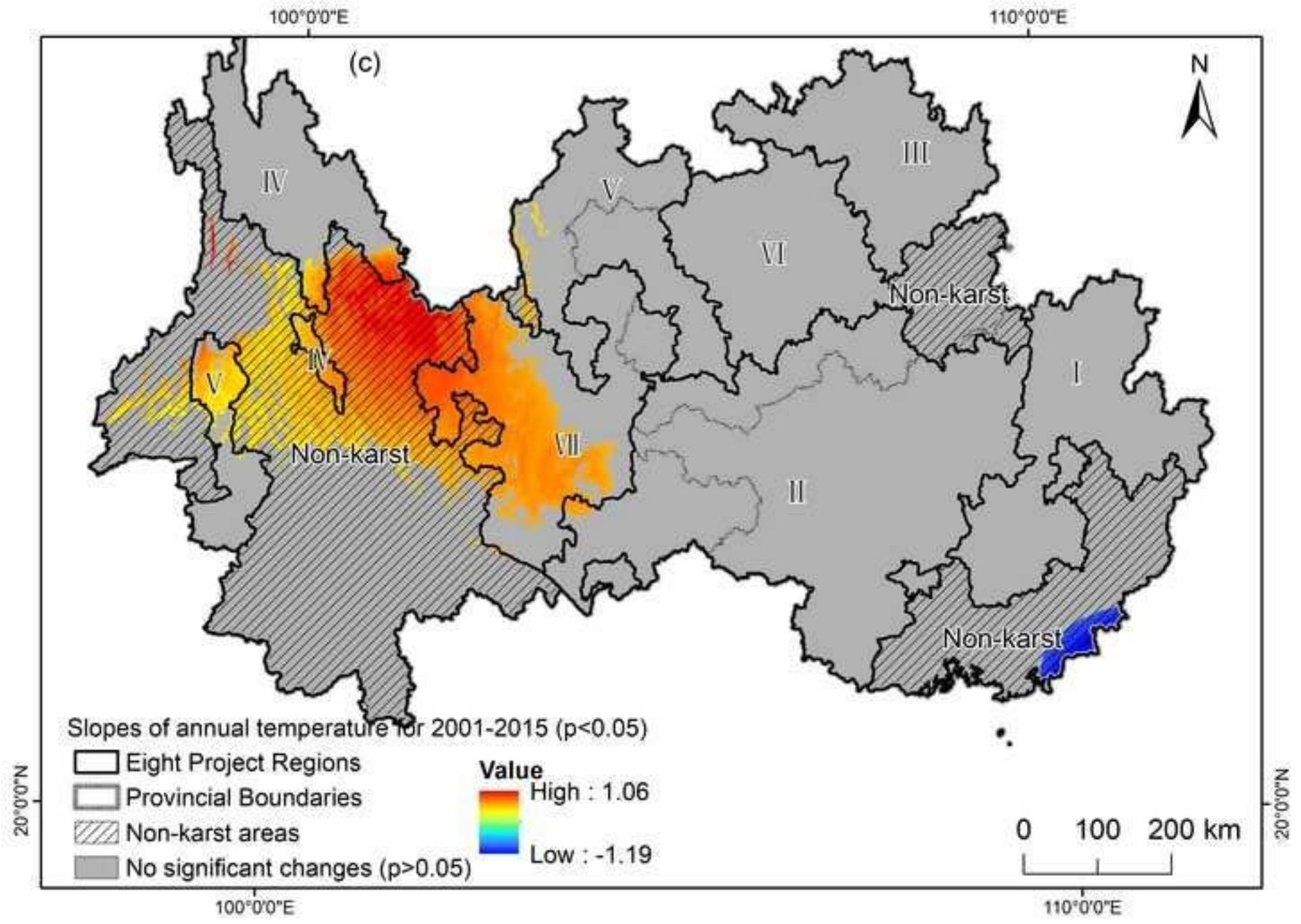

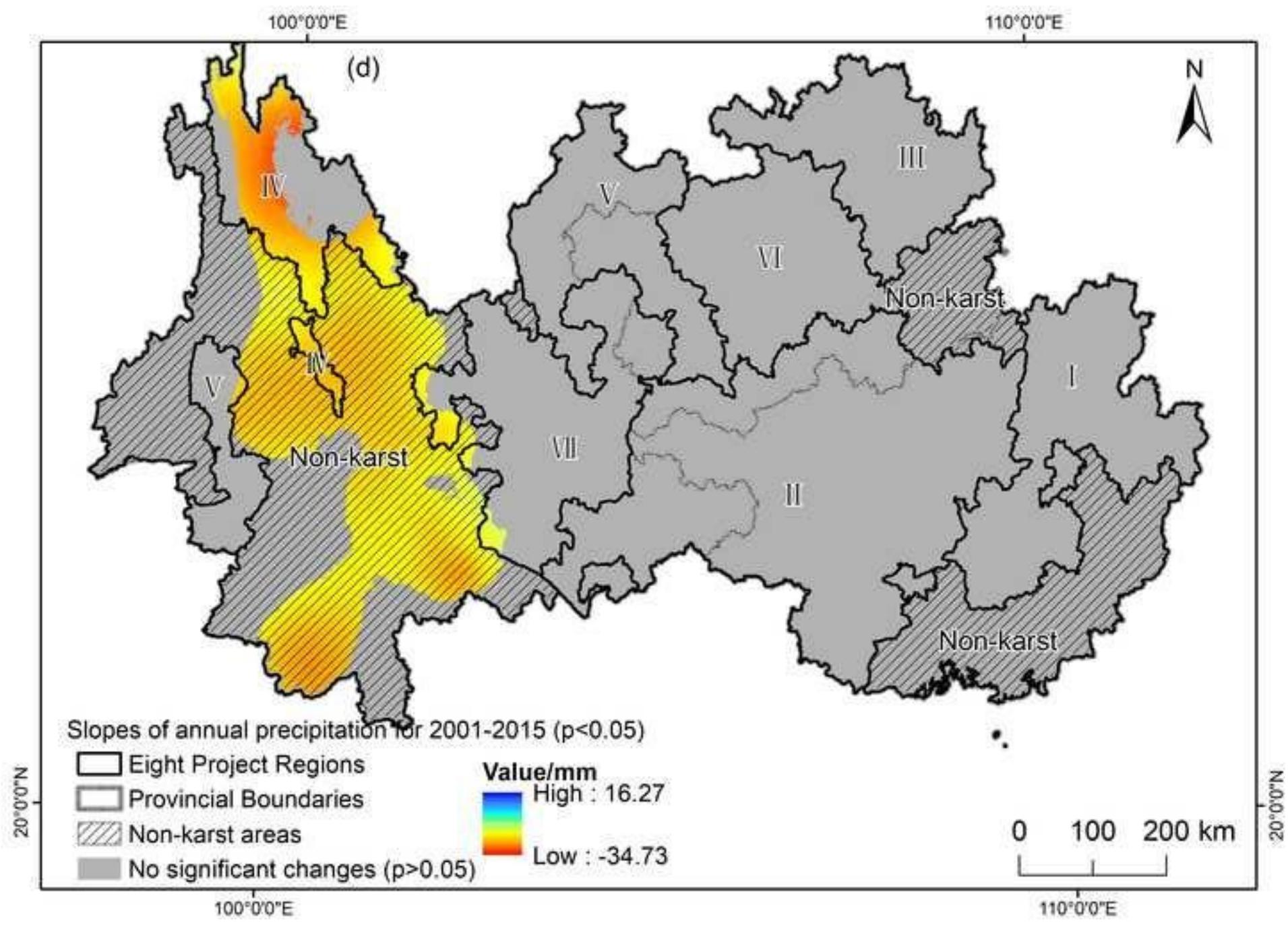

Figure 5Figure 5 Click here to access/download;Figure;fig5_revise.tif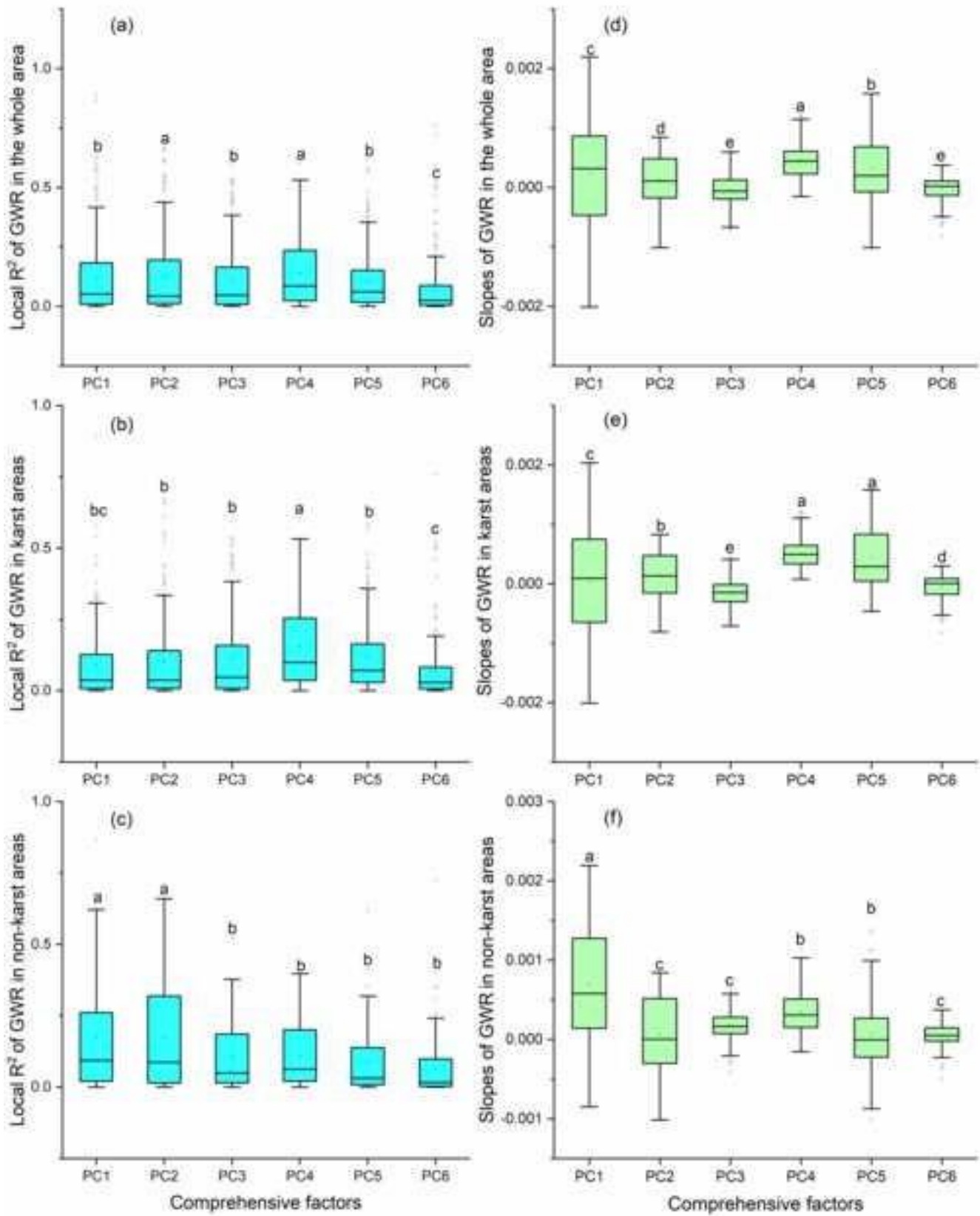

Supplementary material for on-line publication only

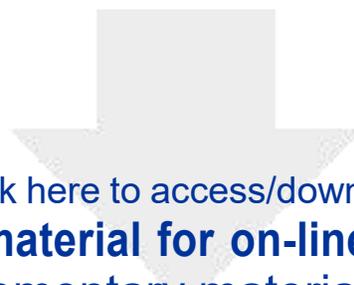

Click here to access/download
**Supplementary material for on-line publication only**
Supplementary materials.docx

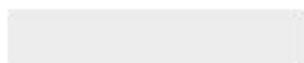

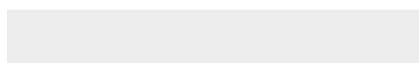

# Credit authorship contribution statement

**Xuemei Zhang:** Conceptualization, Data curation, Methodology, Formal analysis, Investigation, Writing-original draft, Writing-review & editing, Visualition. **Yuemin Yue:** Conceptualization, Supervision, Funding acquisition. **Xiaowei Tong:** Data curation, Writing-review & editing, Funding acquisition. **Kelin Wang:** Conceptualization, Supervision, Funding acquisition. **Xiangkun Qi:** Conceptualization, Data curation. **Chuqiong Deng:** Writing-review & editing. **Martin Brandt:** Conceptualization, Methodology, Writing-review & editing, Visualition.

**Declaration of interests**

The authors declare that they have no known competing financial interests or personal relationships that could have appeared to influence the work reported in this paper.